\documentclass[a4,twocolumn,amsmath,amssymb,nofootinbib,eqsecnum,floatfix,pre]{revtex4}
\usepackage{epsfig}
\psfigdriver{dvips}
\usepackage{graphicx}
\usepackage{color}
\usepackage{dcolumn} 
\usepackage{bm, bbm}      
\usepackage{curves}
\usepackage{epic}
\usepackage{wasysym}
\usepackage{subfigure}
\def\beq{\begin{equation}}
\def\eeq{\end{equation}}
\def\bea{\begin{eqnarray}}
\def\eea{\end{eqnarray}}
\def\nn{\nonumber}
\newcommand{\Tr}{\mathop{\mathrm{Tr}}}

\begin{document}
\title{Criticality of a classical dimer model on the triangular lattice}
\date{\today}
\author{F. Trousselet, P. Pujol, F. Alet and D. Poilblanc}
\affiliation{Laboratoire de Physique Th\'eorique, CNRS, Universit\'e Paul Sabatier, 31062 Toulouse, France.}

\begin{abstract}
We consider a classical interacting dimer model which
interpolates between the square lattice case and the triangular lattice case
by tuning a chemical potential in the diagonal bonds. The interaction energy
simply corresponds to the number of plaquettes with parallel dimers. Using 
transfer matrix calculations, we find in the anisotropic
triangular case a succession of different physical phases as the
interaction strength is increased: a short range disordered liquid
dimer phase at low interactions, then a critical phase similar to the one found for the square lattice, and finally a
transition to an ordered columnar phase for large interactions. The
existence of the critical phase is in contrast with the belief
that criticality for dimer models is ascribed to bipartiteness. For the
isotropic triangular case, we have indications that the system undergoes a 
first order phase transition to an ordered phase, without appearance of an
intermediate critical phase.

\end{abstract}

\maketitle


\section{\label{sec: Introduction}
Introduction}

Dimer models have regained a lot of interest in the
condensed matter physics community thanks to the pioneering work
of Rokhsar and Kivelson (RK)~\cite{Rokhsar1988}. In this original
proposal, dimers represent singlets formed by
pairs of spins $1/2$ in Cuprate materials or frustrated
antiferromagnetic systems. This idea has been followed by many
other proposals giving rise, at low energy, to such kind of
effective models. To cite only a few connections to the physics of
dimers, related again to magnetism, the lowest-energy
configurations of fully frustrated Ising magnets can generally be
mapped onto dimer configurations on the dual
lattice~\cite{Fisher66, MS03}. Other specific (quantum) dimer
models have recently been derived from a spin-orbital model
describing LiNiO$_2$ \cite{vernay}, from the trimerized {\it
kagome} antiferromagnet \cite{zhitomirsky}, or for Heisenberg
antiferromagnets under applied field in the pyrochlore lattice~\cite{Bergman}. Lastly, we note that
hard-core bosons or correlated fermions on frustrated lattices
like the planar pyrochlore lattice can also be mapped onto dimer
representations in the limit of large Coulomb
repulsion~\cite{Senthil,Supersolid}.

Although the RK model is quantum mechanical, there is a
special value of the parameters of the Hamiltonian for which the
ground-state is an equal weight superposition of all dimer tilings of the
square lattice (RK point). Static properties at zero temperature can then be
understood by the purely combinatorial problem of counting dimer tilings of 
the square lattice first solved in the early 60's~\cite{dimer61}. This system
turns out to be a critical model, and more precisely a conformal field theory
with central charge $c=1$. Subsequently to the RK work, Moessner and Sondhi
\cite{MoessSond} have shown that the same dimer model on the
triangular lattice shows, instead of a single point with critical algebraic
correlations, a disordered phase with short range dimer correlations - as
expected from the known physics of its classical counterpart~\cite{FMS02}. From
the interpretation originally given to the dimers, it is legitimate to call
this phase a spin liquid.

More recently, a study performed by Alet {\it et al.}~\cite{Alet0506}  on
a square lattice classical dimer model showed the existence of a
Kosterlitz-Thouless phase transition from a critical phase to a columnar ordered phase for the
dimers. This transition is triggered by including an interaction term,
consisting in the number of plaquettes doubly occupied by parallel dimers.
Subsequently, a quantum mechanical dimer model can be built from this
classical model proving that the single point showing criticality in the
original RK Hamiltonian can be promoted to a whole critical phase~\cite{CCMP}.

An important difference between the square and the triangular lattices for
quantum dimer models lies in the degeneracy of the ground state at the RK
point: it is finite for the triangular lattice with periodic
boundary conditions (PBC) and scales exponentially with the linear size for the square
lattice with the same PBC (see Ref.~\onlinecite{MoessSond} and
Ref.~\onlinecite{Rokhsar1988} respectively for details). This large
degeneracy of the square lattice, related to its bipartiteness, has
been often designated as the responsible of the existence of critical correlations
in this case. It is then natural to investigate the behavior of
interacting dimer model on non-bipartite lattices interpolating between the
square and the triangular lattices, and in particular try to elucidate the
interplay between criticality and bipartiteness. In the non-interacting
case, it is found~\cite{FMS02} that criticality disappears immediately
with the introduction of non-bipartite dimers.

\begin{figure}[!h]
\begin{center}
\includegraphics[width=8cm]{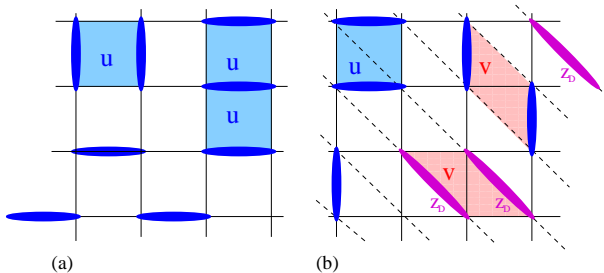}
\caption{\label{2sch} (Color online) Dimer models a) on the square
lattice, with dimer interactions $u$ on square plaquettes; b)
on an anisotropic triangular lattice, with dimer
interactions $u$ and $v$ on square and diamond-like
plaquettes respectively and a fugacity $z_{D}$ for
dimers on diagonal links.}
\end{center}
\end{figure}

In this paper we study an extension of the classical dimer model
with nearest-neighbor interactions already studied
elsewhere~\cite{Alet0506, CCMP} to the case of an anisotropic
triangular lattice. Although a limiting case of this model is the
square lattice, in general the lattice is non-bipartite. In
Sec. II we start by reviewing the results on the square
lattice, and recalling the principles of an associated field
theory used to describe the critical phase. In Sec. III, we
investigate the properties of the anisotropic model that
interpolates between the square and triangular lattices by means
of transfer matrix calculations. There are two independent paths
for the interpolation: the introduction of a fugacity
for \textit{diagonal} bonds but also of the introduction of
interactions between parallel dimers on \textit{diamond-like}
plaquettes (see Fig.~\ref{2sch}). We find for a certain range of
parameters the existence of a critical phase. We determine its
location in the phase diagram as well as transitions from it to
either liquid or ordered phases. In Sec. IV, we find that the
isotropic model on the triangular lattice, does not possess a
critical phase for any interaction strength but could display an
ordered phase for sufficiently large interactions. Finally,
Sec. V contains a discussion of the various results and
conclusions.

\parskip 10pt


\section{\label{revsqr}
Criticality of the dimer model on the bipartite square lattice}

Let us start by reviewing the main results of the square lattice
dimer model. In this paper, we will only consider fully-packed
dimers with a hard-core constraint: there must be one and only
one dimer coming out of each site. The non-interacting model can
be solved exactly on the square lattice by expressing the
partition function of the system as a Pfaffian, which enables to
obtain analytic expressions for the free energy and correlators,
on either finite or infinite systems \cite{dimer61,Fish-Steph}. On
the square lattice, dimers are found to be in a critical phase:
dimer-dimer correlations decay algebraically as $1/r^2$ with
relative distance $r$, whereas the monomer-monomer (defining a
monomer as a site with no dimer) correlations behave as
$1/\sqrt{r}$. Classical dimers on the square lattice have also
been studied within a finite-temperature model with nearest neighbors
interactions, {\it i.e.} an interaction $-u$ between parallel dimers on
the same plaquette ($u>0$ corresponding to attractive
interactions)~\cite{Alet0506}. Here the model is not integrable
anymore and it was found numerically that the
system is critical down to a finite temperature $u_{\rm col}/T=1.54(3)$, 
where it gives rise to a low-temperature columnar
phase with dimers aligned in columns. Monomer-monomer correlations
decay to $0$ at long distances, either algebraically (in the
critical phase) or exponentially (in the columnar phase). From now
on, unless it is specified otherwise, we set $T=1$
and let the different dimer-dimer interactions ($u$ 
for the purely square lattice) vary. Nevertheless, we still refer to
\textit{high-T} (respectively \textit{low-T}) regions characterized by
small (resp. large) values of the involved couplings.

The criticality encountered in the high-T phase can be understood
in the framework of a height field theory. For each dimer
configuration on the square lattice, a scalar height field with
quantized values (multiples of $\frac{1}{4}$) can be defined
microscopically on the dual lattice~\cite{Blote}. The spatial
variations of the height field between neighboring sites of the
dual lattice are entirely determined by the presence of dimers
between them, thanks to the lattice bipartiteness. The long-wavelength modes of this height correspond to a
coarse-grained height field $\Phi({\bf r})$ defined in continuum space,
and the physics of the model is captured by the
action~\cite{Alet0506,Henley}: 
\begin{equation} 
S[\Phi]=\int d^{2} {\mathbf r} g \pi |\nabla \Phi({\mathbf r})|^{2} + V
\cos(2\pi p \Phi({\mathbf r}))
\label{Action1} 
\end{equation}
The cosine term of this action is a
\textit{locking potential} that favors $p$ \textit{flat}
configurations ($p=4$ for the square lattice), corresponding
microscopically to the columnar configurations. The
$|\nabla \Phi|^2$ term, corresponding to the cost of fluctuations
around these flat configurations, account for the entropy of dimer
coverings. In the non-interacting case, the value of the stiffness
is fixed by the exact results of Ref.~\cite{dimer61,Fish-Steph} to
be $g=\frac{1}{2}$  (see also Ref.~\cite{Henley97}).  The
renormalization of the stiffness constant accounts for the role of
interaction: $g$ increases with $u$.

As long as the cosine operator in Eq.~(\ref{Action1}) is irrelevant in the
renormalization group sense, this action defines a conformal field
theory with central
charge $c=1$ and the system is critical. The dimer operator (which gives the
local dimer density) is composed of a $\cos(2\pi\Phi)$ and gradient
terms~\cite{FradHuse04}, and has a scaling dimension $d_{1,0}=\frac{1}{2g}$ ,
corresponding to half of the exponent of dimer-dimer
correlations. Similarly, the locking potential has a dimension
$d_{p,0}=\frac{p^2}{2g}=\frac{8}{g}$, and becomes relevant when $g \ge 4$.
Within the standard  Coulomb gas description~\cite{CG} of this theory,
we can define operators corresponding to the
insertion of a particle of electric and magnetic charges
$e$ and $m$ respectively. In general, the dimension of such an electromagnetic
vertex operator is $d_{e,m}=\frac{e^2}{2g} + \frac{g m^2}{2}$. In the height
model discussed here, the operator
with $(e,m)=(1,0)$ is the dimer operator already mentioned; similarly, the
monomer operator corresponds to inserting a magnetic particle with
$(e,m)=(0,\pm 1)$ (the sign of the charge depends on the sublattice where
the monomer is inserted).
In the critical phase, dimer-dimer and
monomer-monomer correlators decay as power laws of the distance and the
decay exponents are $2d_{1,0}$ and $2d_{0,1}$ respectively.
Going back to the height picture, the insertion of magnetic charges (for instance monomers
or links breaking the bipartiteness) corresponds to dislocations
and, as we show below, can be treated as perturbations in the coarse-grained 
field theory.

Prior to the presentation of our strategy to attack the
dimer problem, we first introduce the numerical techniques used here and
illustrate them on the example of the square lattice interacting dimer model.
In this study we use a Transfer Matrix (TM) approach to determine the domains
of existence of the different phases of the dimer model,
computed on a torus of longitudinal and transverse sizes
$L_{\tau}$ and $L$ respectively. The TM techniques allow
to compute exponents of critical dimer-dimer and monomer-monomer
correlations and also correlations themselves (see
Appendix), on systems with $L_{\tau}$ large enough to be considered infinite.

For a critical phase such as the one encountered in the high-T region of the
square lattice dimer model, we can compute the decay exponents of
correlation functions in two ways: either by a direct inspection of the
real-space decay of correlation functions (see the description in
Sec.\ref{corr} of
Appendix), or by the leading eigenvalues of the TM (see Sec.\ref{CFT} of
Appendix). Both methods are guided by a conformal field theory (CFT) analysis of the critical phase,
and the latter is more precise to determine exponents as it allows to use
efficiently topological and translation symmetries of the TM
(see Sec.\ref{CFT} in Appendix). The calculations of correlation functions
in real space offer however the advantage of providing more physical
informations (for example about the correlation length) when the system is not
critical.
Another important quantity, namely the central charge $c$ of the CFT, can
also be extracted from the finite-size scaling of the TM largest eigenvalue
when the system is critical (see Sec.\ref{CFT} of Appendix).

\begin{figure}[!h]
\begin{center}
\includegraphics[width=8cm]{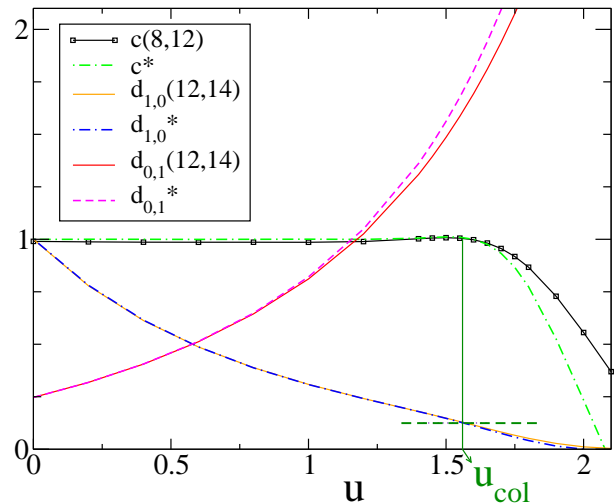}
\caption{\label{plat0} (Color online) Estimates and
extrapolations from different estimates (see Appendix for details and
notations) of the central charge $c$ and of the scaling dimensions $d_{1,0}$ and
$d_{0,1}$ in the dimer model on the square lattice as a function
of interaction strength $u$.}
\end{center}
\end{figure}

We illustrate the validity of this numerical approach in Fig.~\ref{plat0},
where the critical phase is evidenced by the $c=1$ plateau of the
central charge (estimated using TM largest eigenvalues for system widths up
to $L=14$). The scaling dimensions of dimer $d_{1,0}$ and monomer $d_{0,1}$
operators (obtained from subleading eigenvalues) are also displayed
according to the discussion above,
the transition to the columnar phase is characterized by a value $g=4$ of the
stiffness, thus a value $d_{1,0}=1/8$. This criterion is used to estimate the
transition temperature (estimates of $d_{0,1}$ can also be used in principle,
 but they are more sensitive to finite-size effects in that temperature
 range). The results of Fig.~\ref{plat0} lead to an estimate 
$u_{\rm col}=1.563(7)$, 
in good agreement with previous results~\cite{Alet0506,CCMP}.

\section{\label{iplat}
CRITICAL PHASE IN A DIMER MODEL INTERPOLATING THE SQUARE AND TRIANGULAR
LATTICES}

In this section, we investigate whether the bipartiteness
of the square lattice is a condition for the existence of the critical
phase, by building a model defined on a lattice interpolating continuously
between the square and the triangular lattices.

\begin{figure}[!h]
\begin{center}
\includegraphics[height=5cm]{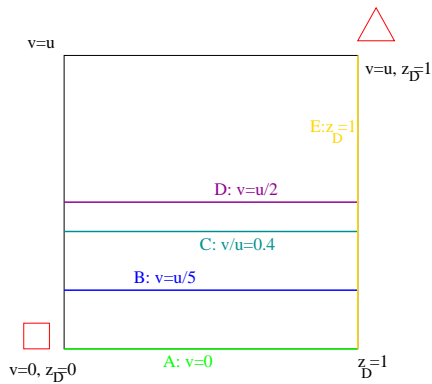}
\caption{\label{path} (Color online) Different parameter-space paths (A to E)
taken in the TM calculations in section~\ref{iplat}.}
\end{center}
\end{figure}

{\it Definition of a model with lattice and interaction anisotropies --- }
For  commodity, we represent the triangular lattice as a deformed square
lattice with bonds in the $x$, $\tau$ and $\tau-x$ direction (see Fig. 
\ref{2sch} and Appendix). The interpolation between both lattices is made by 
assigning a fugacity $z_{D}$ to the diagonal bonds. Since the elementary
plaquettes of the triangular lattice are more numerous than those of the
square, we need to define another parameter $v$ characterizing the
interaction on \textit{diamond-like} plaquettes ({\it i.e.} non-square
four-site plaquettes). We finally keep the notation $u$ for the interaction
between parallel dimers on a square plaquette. These elementary energy scales
in the problem are
illustrated in Fig. \ref{2sch}. For $z_{D}=0$ and $v=0$, we have the
classical dimer model on the square lattice discussed in Sec.~\ref{revsqr},
while the pure triangular dimer model (that will be discussed in
Sec.~\ref{isotri}) corresponds to both $z_{D}=1$ and $v/u=1$. To describe the 
system in function of these two anisotropy parameters, we take
the approach of varying either $z_{D}$ or $v/u$ while keeping the other
parameter constant. The different paths in the $(z_{D},v/u)$ space where 
phase diagrams were computed numerically are represented in Fig.~\ref{path}.

{\it Description of expected phases and transitions--- }
In the non-interacting limit ($u=v=0$) of the present model, the system is
disordered with exponentially decaying correlations for any fugacity
$z_{D}>0$: criticality is destroyed by an infinitesimal proportion of
diagonal bonds~\cite{FMS02}.
This can be accounted for in the height field theory by adding a relevant term
in the action Eq.~(\ref{Action1}), which can be understood by noticing that
adding a single diagonal dimer on the square lattice corresponds to
inserting two monomers on the same sublattice, {\it i.e.} two magnetic
charges of the same sign~\cite{CCMP2}. The corresponding $m=2$ 
vertex operator
is written in terms of the field $\Theta$, dual to the height field $\Phi$,
and leads to the new effective action:
\beq
S[\Phi]=\int d^{2}{\bf r} \pi g |\nabla \Phi ({\bf r})|^{2}
+ V \cos(8 \pi \Phi({\bf r})) - \lambda \cos(2 \Theta({\bf r}))
\label{Action2}
\eeq

In the absence of interactions, this new perturbing field (the $\cos(2\Theta)$ term)
corresponds to the mass term of the free Majorana doublet arising in the large scale regime
within the Pfaffian description of the dimer model \cite{FMS02}. 
For a generic value of the stiffness, the scaling dimensions of the two perturbing terms are
$d_{4,0}=\frac{8}{g}$ and $d_{0,2}=2g$ respectively. The fact that the
introduction of diagonal bonds is indeed 
relevant in the non-interacting limit is seen from the value of the 
stiffness 
in this case $g=\frac{1}{2}$ (since $d_{0,2}=1<2$). However, the key point 
of our study is that the $\cos(2\Theta)$ term becomes irrelevant for $g>1$ 
whereas the locking potential is relevant only for $g\geq 4$. As we can tune 
the stiffness constant by modifying the interaction strength
between parallel dimers ($u$ and $v$ terms), we are hoping to reach the
$1 < g < 4$ window, {\it i.e.} a critical phase on the triangular lattice,
within the interacting dimer
model defined above. The predicted phase diagram is therefore as follows:
first, a high-$T$ liquid phase (for $1/2<g<1$ in the unperturbed model), then
a transition to a critical phase ($1 < g < 4$) at intermediate couplings and
finally a low-$T$
columnar ordered phase (for $g>4$). The dimer $d_{1,0}$ and monomer
$d_{0,1}$ scaling exponents are predicted to be both equal to $1/2$ at the
transition from the liquid to the critical phase, and $1/8$ and $2$
respectively at the critical-columnar transition. In both cases, transitions
out of the critical phase are expected to be of Kosterlitz-Thouless type. This 
analysis holds only in the perturbative regime for the diagonal
dimer fugacity $z_{D}\ll 1$, and we now address the question whether this
scenario is realized for an arbitrary lattice anisotropy $0<z_{D}\leq 1$ by
means of numerical TM calculations.

\begin{figure}[!h]
\begin{center}
\includegraphics[width=7cm]{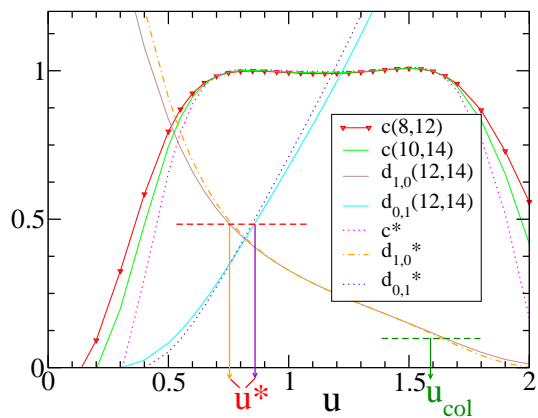}
\caption{\label{plat1} (Color online) Central charge $c$, estimates and
extrapolations of the dimer exponent $d_{1,0}$ and monomer
exponent $d_{0,1}$, as a function of $u$ for a diagonal dimer
fugacity $z_{D}=0.4$ and $v=0$.}
\end{center}
\end{figure}

{\it Perturbation of the square lattice model by diagonal bonds --- }
We first consider the model with no additional interactions $v=0$ and turn on
the diagonal dimer fugacity (path A in Fig.~\ref{path}). Our numerical estimates of the central charge $c$ and
 exponents $d_{1,0}$ and $d_{0,1}$ are displayed in Fig.~\ref{plat1} as a function of the
coupling strength $u$ for the specific value $z_{D}=0.4$. We clearly observe
the emergence of a $c=1$ plateau witnessing a critical phase, for a wide
range of couplings. The points $u_{1}$ and $u_{2}$ where $d_{1,0}=1/2$ and 
$d_{0,1}=1/2$ 
correspond roughly to the high-T limit of the $c=1$ plateau, as predicted 
above; from these criteria, the transition between the critical and liquid 
phase is located at $u^*=(u_{1}+u_{2})/2=0.80(5)$, the error bar being 
estimated by $(u_{2}-u_{1})/2$. Similarly, the low-T end of the plateau at 
$u_{col}=1.566(8)$, corresponding 
to the entrance into the columnar phase, coincides with the criterion
$d_{1,0}=1/8$, confirming the analysis above. We use estimates of $d_{1,0}$
to locate the columnar transition, since as on the square lattice estimates 
of the monomer exponent 
$d_{0,1}$ are more affected by finite-size effects.

\begin{figure}[!h]
\begin{center}
\includegraphics[width=7.5cm]{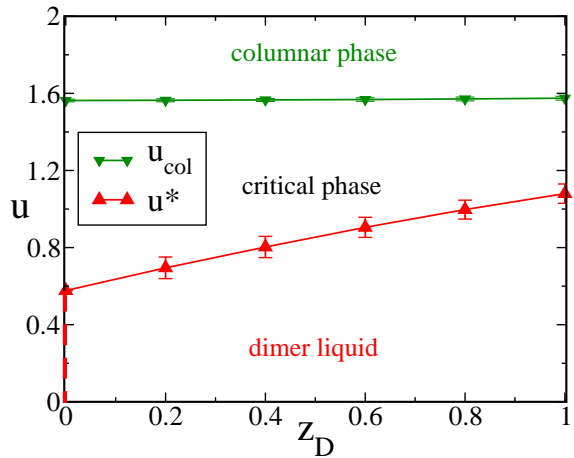}
\caption{\label{diag1} (Color online) Phase diagram for $v=0$ (line A in 
Fig.~\ref{path})
  as a function of diagonal dimer fugacity $z_{D}$ and square-plaquettes
interaction $u$. Note that on the $z_D=0$ line, the small $u$
region also  corresponds to a critical phase (dashed line).}
\end{center}
\end{figure}

Repeating the same analysis for different values of the diagonal dimer
fugacity, we obtain, as a function of $z_{D}$ and $u$, the phase diagram of
Fig.~\ref{diag1}. The main features are:
{\it (i)} the transition temperature to the columnar phase is essentially
not affected by the diagonal dimer fugacity and remains close to the value
obtained in the square lattice; {\it (ii)} as found in
Ref.~\onlinecite{FMS02}, the system with no interaction $u=0$ is a gapped
liquid irrespective of the diagonal dimer fugacity; {\it (iii)} the dimer model
(with $v=0$) comprises a critical phase even on the isotropic
triangular lattice ($z_{D}=1$), where it extends from $u^{*}=1.08(5)$ up to
$u_{col}=1.575(10)$.

These results are in full agreement with the predictions of the
field-theoretical analysis developed above.  
We emphasize also here that the lattice bipartiteness is
not a necessary condition to have criticality in a dimer model,
and that this condition should rather be replaced by a condition
on the existence of a stiffness constant window $1<g<4$ in an
unperturbed model.

{\it Introduction of a finite interaction anisotropy $v/u$ --- }
We now introduce the interaction term $v$ on the diamond-like
plaquettes. The $u$ and $v$ terms are in competition, since they
tend to favor different dimer orderings. When $v < u$, the square
columnar configurations will eventually dominate at low enough
$T$, but the presence of the frustrating  $v$ interactions will
shift down the transition temperature to the ordered phase. As the
algebraic correlations of the critical phase also correspond to a
square-like columnar ordering at lower temperatures, we also
expect the extent of the critical phase to shrink down as the
strength of the $v$ interactions is increased. From the point of
view of the effective action Eq.~(\ref{Action2}), we expect $g$ to
{\it decrease} with the $v$ frustrating interactions.

\begin{figure}[!h]
\begin{center}
\includegraphics[width=7cm]{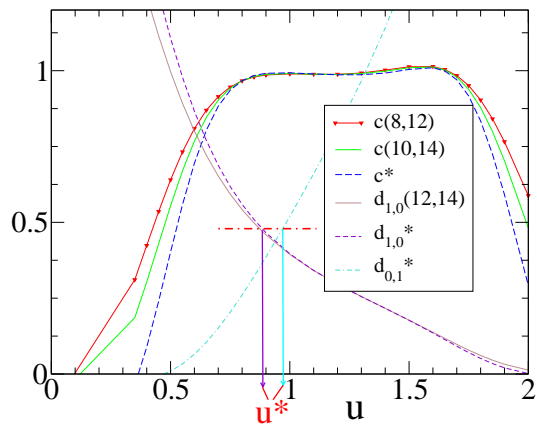}
\caption{\label{plat2} (Color online)  Central charge and exponents (estimates and
extrapolations) for an interaction anisotropy $v/u=0.2$ and
diagonal dimer fugacity $z_D=0.4$.}
\end{center}
\end{figure}

For a sufficiently small $v/u$ parameter, the behavior of the
system is found to be identical to the case $v=0$, with a critical
phase, evidenced by a $c=1$ plateau between the dimer liquid and
the ordered phase. This is illustrated in Fig.~\ref{plat2} for
anisotropy parameters ($z_{D},v/u$)=($0.4,0.2$). As expected, the
transition towards the ordered phase is slightly shifted towards
lower temperatures ($u_{\rm col}=1.624(8)$ for $v/u=0.2$ while
$u_{\rm col}=1.566(8)$ for $v=0$). The high-T boundary of the
critical phase is more strongly affected by the presence of
$v$ interactions: for the same lattice anisotropy, $u^{*}=0.92(6)$
for $v/u=0.2$ to be compared with $0.80(5)$ for $v=0$.

\begin{figure}
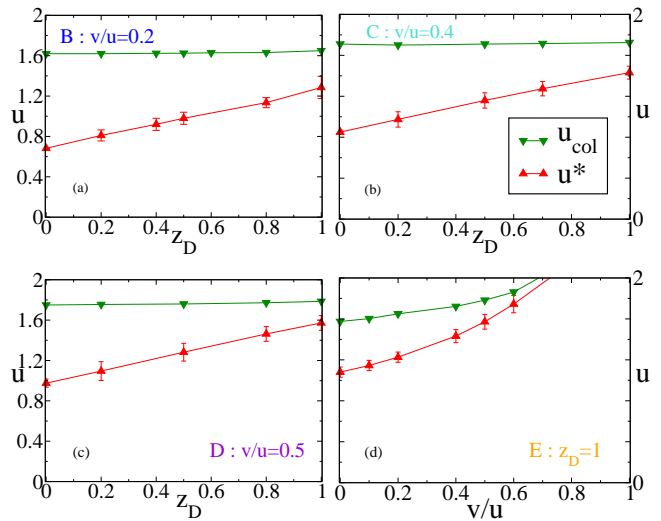

\begin{center}
\includegraphics[width=4.2cm]{diaguDbyd10_p02.eps}
\includegraphics[width=4.2cm]{diag_p04_73.eps}\\
\parskip 5pt
\includegraphics[width=4.2cm]{diaguDp05.eps}
\includegraphics[width=4.2cm]{diaguv_D1.eps}
\caption{\label{diag2} (Color online)  Phase boundaries $u^{*}$ and 
$u_{\rm col}$ as a function
  of either $z_{D}$ [(a), (b) and (c)] or $v/u$ [(d)], corresponding
  respectively to the paths B, C, D and E in Fig.~\ref{path}. Error bars
on $u_{\rm col}$ are smaller than symbol size.}
\end{center}
\end{figure}

Using the same procedure as before, the transition points $u^{*}$ and
$u_{\rm col}$ were determined in the regions of the phase diagram corresponding
to paths B,C,D (with respectively $v/u=0.2,0.4$ and $0.5$), and E
(for $z_{D}=1$) and the corresponding phase diagrams are reported
in Fig.~\ref{diag2}. As anticipated, we clearly see that the effect of adding the
interaction $v$ is essentially to shift both boundaries of the
critical phase towards lower temperatures: in other words, the
frustration due to the $v$ terms has a net tendency to stabilize
the dimer liquid and to destabilize the ordered (columnar) or
quasi-ordered (critical) phase. For a given fugacity $z_{D}$, the
extent of the critical phase indeed decreases as $v/u$ increases.
The transition between the dimer liquid and the critical phase
turns out to be more affected by $v/u$ than the transition towards
the columnar phase. We can interpret this with the following rough
argument: $v$ interactions effectively increase the weight of
diagonal bonds in the dominant configurations, enhancing the
perturbation caused by the diagonal bonds and therefore decreasing
the extent of the critical phase from the high-$T$ direction.

For large values of $v/u$, a fair determination of phase
boundaries becomes more difficult as finite-size effects are getting more
important. For instance, the determination of $u^{*}$ with the dimer and
monomer exponents criteria give values increasingly far from each
other, resulting in larger error bars in the phase diagrams of
Fig.~\ref{diag2}c and d. In practice for $v/u > 0.5$, both our numerical
results and the way of analyzing them break down due to uncontrolled finite
size effects. This is actually not a surprise, as the CFT-guided analysis of
the TM results (see Sec. \ref{CFT} in Appendix) relies on the existence of an
unperturbed critical window in the model, which is no longer present for large
$v$ interactions. For $v/u>0.5$, we also find that the second largest
TM eigenvalue is no more found in the $q=\pi$ symmetry sector in
the low-T phase. This indicates a change in nature of the lowest-energy
excitations and that the scheme along which criticality was previously 
understood is not valid anymore. From these arguments and even if our finite-size
results become less reliable, it is clear that they are no
longer compatible with the existence of an extended critical phase
for $v/u > 0.5$: this is exemplified by the absence of a plateau
in the numerical estimate of the central charge. Finally, we note that in parallel to the shrinking of the critical phase
observed when increasing $v/u$, the columnar transition seems to be
progressively shifted towards the $T=0$ limit as the interaction anisotropy
parameter approaches $1$ (see Fig.~\ref{diag2}d). This is again in
agreement with our previous qualitative arguments on the effect of $v$
interactions.


\section{\label{isotri}
Dimer ordering on the isotropic triangular lattice}

We now turn to the more complex situation of the isotropic
triangular model, where $z_D=1$ and $v=u$ (in this section, the
isotropic interaction will only be denoted by $u$). The high-$T$
limit is in this case well understood: the system is in a dimer
liquid phase, with exponentially decaying dimer-dimer and
monomer-monomer correlation functions. This has been shown exactly
for $u=v=0$ in Ref.~\onlinecite{FMS02}, and we could confirm it
for small values of the interaction strength. For intermediate
$u=v$ values, as expected from the results of Sec.~\ref{iplat}
for $v/u>0.5$, we find no evidence for a critical phase in our TM
calculations (in particular, no $c=1$ plateau).

\begin{figure}[!h]
\begin{center}
\includegraphics[width=8cm]{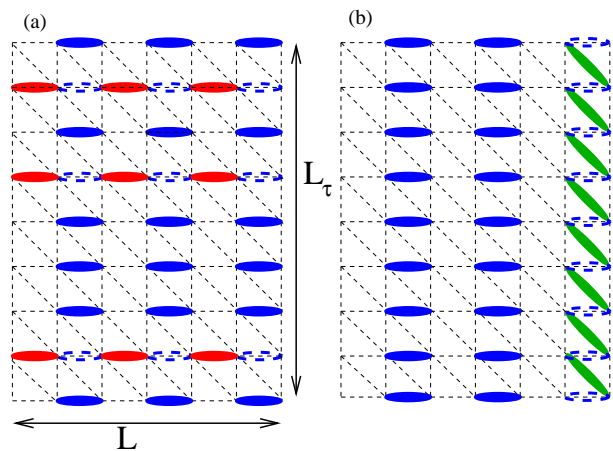}
\caption{\label{def} (Color online) Different types of zero-energy defects on the
(isotropic) triangular lattice dimer model, from a specific columnar
ordered ground-state: (a) \textit{line-shifting modes}; (b) \textit{column-flipping modes}.}
\end{center}
\end{figure}

Whereas we always found a low-T columnar ordering up to now, the
low-$T$ behavior of the isotropic triangular model reveals more
complex. Indeed, from the isotropy of the interactions and since
all dimer fugacities are equal, it appears that configurations that minimize
the energy are more numerous. At first glance, $12(=6\times 2)$ columnar 
ordered ground-state
configurations emerge ($6$ for the number of ways to put a dimer
on a link connected to a given site and $2$ for the possible directions 
of a dimer column, given an orientation of dimers). However,
as first remarked in Ref.~\onlinecite{MoessSond}, the number of
configurations that minimize the energy is much larger as one can
easily create zero-energy defects by just translating a line of
dimers or flipping all dimers along a column. These two types of
moves from a given columnar ground-state are represented in
Fig.~\ref{def}. The almost extensive ground-state degeneracy
generated by these costless defects clearly changes the picture
for the low-$T$ behavior of the isotropic triangular dimer model.
Indeed, at $T=0$, strictly speaking, the system is no longer
ordered with respect to a local order parameter such as the
columnar one defined for the square lattice. However, it is
still possible that at finite $T$, the thermal fluctuations select a specific 
ordering pattern: this would be an illustration
of the ``order by disorder'' effect~\cite{Villain,SH}.

In fact, for the quantum dimer model on the triangular lattice,
Moessner and Sondhi found in perturbation theory (with respect to
the quantum kinetic terms) that indeed (quantum) fluctuations
select the $12$ ordered columnar states~\cite{MoessSond}. Noticing that in our
classical model, {\it thermal} fluctuations play the same role as
{\it quantum kinetic} fluctuations in the quantum dimer
model~\cite{Alet0506} since both essentially count the number of
flippable plaquettes in a given state, it is likely that thermal fluctuations
(instead of quantum fluctuations) trigger a similar
order-by-disorder scenario here. From general
arguments~\cite{SH}, we expect that if it is indeed the case, the
transition temperature to the columnar phase should be quite low. It could also
happen that, for our specific model, this transition never occurs
and the systems stays in a liquid phase down to $T=0$.

\begin{figure}[!h]
\begin{center}
\includegraphics[width=8cm]{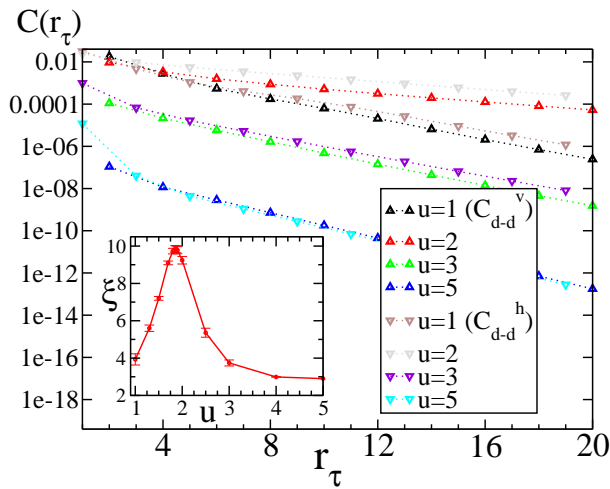}
\caption{\label{fig4} (Color online) Connected dimer-dimer correlations as a
function of the distance $r_{\tau}$ along the main axis of a torus of
dimensions $(L_{\tau},L)=(60,6)$, for interaction
strengths $u$ between 1 and 5. Correlations of vertical
(respectively horizontal) dimers are represented by up (resp.
down) triangles. For clarity, we only displayed horizontal (resp.
vertical) dimer correlations at odd (resp. even) distances. Inset:
estimated correlation length $\xi$ as a function of $u$.}
\end{center}
\end{figure}

We now try to settle this issue with the help of numerical TM
calculations. As the critical phase disappeared, we can no longer use the
CFT-based analysis of the previous sections and we have to resort to more
standard thermodynamical means of detecting the hypothetic transition to a
low-$T$ ordered phase. We first computed dimer-dimer correlations by TM
iterations (see Appendix). In Fig.~\ref{fig4}, connected
correlation functions are displayed as a function of dimer-dimer
distance $r_{\tau}$ for the example of a $(L_{\tau},L)=(60,6)$ torus for 
interaction strength $u$ up to 5. As exemplified by the log-linear scale, 
they are clearly short-ranged with
an exponential decay (plus an oscillating part depending on the parity of
$r_{\tau}$). We obtain an estimate of the correlation length $\xi$ by fitting 
the correlations to an exponential decay $\exp(-r_{\tau}/\xi)$. The resulting 
curve $\xi(u)$, displayed in the inset of Fig.~\ref{fig4},
shows a well-pronounced peak at a finite value of $u$, which could
correspond to an ordering temperature. We also find that the peak positioned at
$u=1.9(1)$ for the system size $L=6$, shifts towards higher interaction
strength $u=2.22(5)$ for $L=8$.

Another thermodynamical insight is given by the specific heat per
site $c_{v}=\frac{1}{L.L_\tau}d\langle E \rangle /dT$ , which is
also accessible to the TM calculations (see Sec.~\ref{part} of
Appendix for details). Fig.~\ref{fig6} displays the specific heat
as a function of $u$ for different system widths $L$. For each
value of $L$, the specific heat peaks at a temperature close to
the ones of the $\xi(u)$ curves. As $L$ increases, the peak in
$c_{v}(u)$ sharpens and shifts to lower temperatures. We discuss
below the finite-size scaling of the position of the peak but
already note at this stage that the clear narrowing of the
specific heat peak with increasing system size, conjugated with
the absence of a power-law envelop, is very suggestive of a {\it
first-order} phase transition.

\begin{figure}[!h]
\begin{center}
\includegraphics[width=7cm]{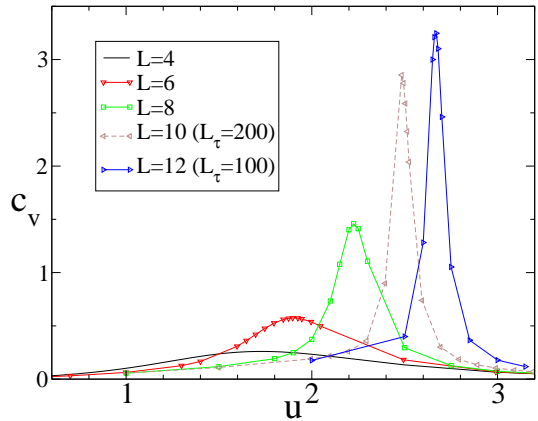}
\caption{\label{fig6} (Color online) Specific heat per site 
$c_{v}=\frac{1}{L.L_\tau} \frac{d\langle E \rangle}{d T}$ as a
function of $u$ for systems of different widths (see Appendix for
details of computation).}
\end{center}
\end{figure}

Having possible signs of a phase transition, we now try to find an order
parameter for the low-$T$ phase. As already discussed, this a not a simple
task because of the zero-energy modes that are responsible for the large
ground-state degeneracy. The specific geometry of the TM is of help
here to characterize the appearance of long-range order. Consider a very
long cylinder $L_{\tau}\gg L\gg 1$ with PBC in the small $L$ direction. With
this geometry, the ground-states having dimers perpendicular to the
long direction (horizontal dimers)  are much more numerous than ground-states
with only vertical or diagonal dimers. Indeed, there are $2^{L_{\tau}}$
horizontal ground-states (corresponding to the \textit{line-shifting modes},
see Fig.~\ref{def}a). The same line-shifting modes provide only a much
smaller ($2^L$) degeneracy for vertical or diagonal ground-states. 
The other family of low-energy modes (\textit{column-flipping modes}) give a
small $O(2^{L/2})$ degeneracy for the horizontal ground-states (see Fig.~\ref{def}b), and a
$O(2^{L_{\tau}/2})$  degeneracy for both vertical and diagonal
ground-states. Consequently, in the limit of large $L_\tau$, we expect
horizontal line-shifting modes to predominate. 

\begin{figure}[!h]
\begin{center}
\includegraphics[width=\columnwidth]{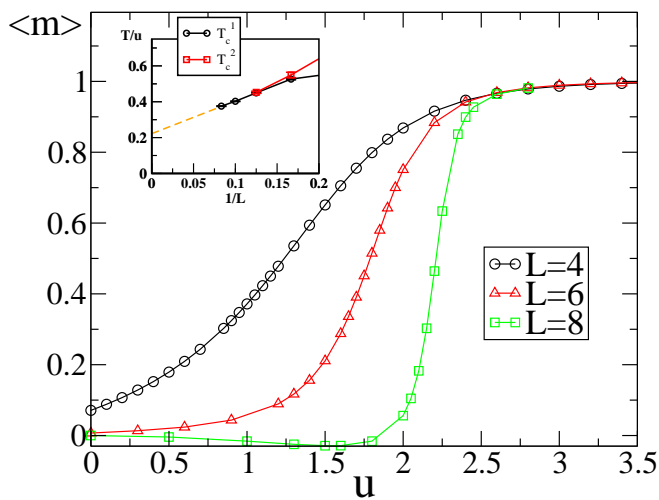}
\caption{\label{m1m2} (Color online) Proposed order parameter $\langle m \rangle$ for the
  long cylinder geometry (see text) versus coupling strength $u$ for 
different lattice sizes. Inset: finite-size transition temperatures
  $T_{c}^1$ and $T_{c}^2$ (maxima of $c_{v}$ and $d
  \langle m \rangle/d u$ respectively) versus $1/L$.}
\end{center}
\end{figure}

If there is long-range order in the system, the previous analysis
indicates that the long cylinder geometry induces a preferential
ordering in the horizontal dimers (for entropic reasons).
Consequently, we consider the average occupation of horizontal
bonds on a given line of the lattice as a good indicator of a
possible phase transition. At infinite temperature,
occupations of all bonds are all likely and equal to $1/6$. In
contrast, from the arguments above, 
at very low $T$, the probability for a horizontal bond
to be occupied on an infinitely long cylinder is $P_{-}=1/2$. 
Defining $$\langle m \rangle=\frac{P_{-}-1/6}{1/2-1/6}=3P_{-}-\frac{1}{2},$$
we expect the ``order parameter'' $\langle m \rangle$ to vanish in
the low-coupling limit $u=0$ and to saturate to 1 at large enough
coupling. Even if, strictly speaking, this above argument is not rigorous (because $L_{\tau}$ and $L$ are both finite in our
computations), we expect $\langle m \rangle$ to reflect a true
physical behavior. $\langle m \rangle$ is shown in Fig.~\ref{m1m2}
for lattice widths $L=4,6,8$ and we clearly observe the predicted
behavior (saturation to $0$ at small $u$ and to $1$ at large
couplings). As for the specific heat data, we observe that the
shape of $\langle m \rangle$ is strongly influenced by the system
size, with similar trends: as $L$ increases, the temperature
region of ordering (where $d \langle m\rangle / du$ is maximal) is
getting narrower and is shifted towards lower temperatures.

\begin{figure}[!h]
\begin{center}
\includegraphics[width=7cm]{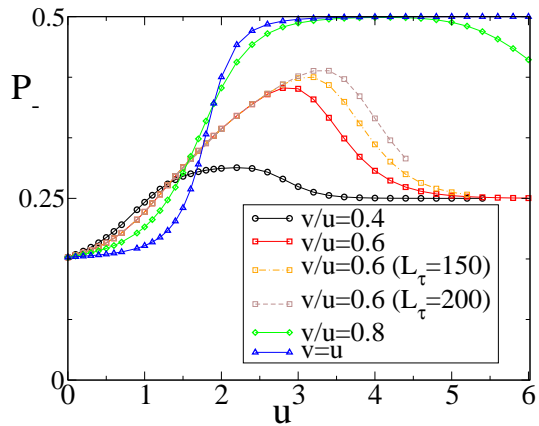}
\caption{\label{manis}(Color online) Average occupation $P_{-}$ of horizontal 
bonds computed on a row of an isotropic triangular lattice of dimensions
$L=6$ and $L_{\tau}=100$ (if not specified), for different interaction 
anisotropy parameters
$\frac{v}{u}=0.4;0.6;0.8;1$, as a function of interaction strength.}
\end{center}
\end{figure}

In order to have a better understanding of this ordering phenomenon, we have
computed the order parameter $\langle m \rangle$ away from the
purely isotropic case, for a diagonal dimer fugacity $z_{D}=1$ but
for values of $v/u$ lower than $1$. The evolution of
$\langle m \rangle$ with $u$ is shown on
Fig.\ref{manis} for different interaction anisotropies $v/u
=0.4;0.6;0.8;1$. As soon as $u$ is larger than $v$, the mean
occupation of horizontal links $P_{-}$ is $1/4$ at zero
temperature, which is consistent with the fact that, as in the
purely square columnar state, only $4$ configurations have the
minimal energy. Nevertheless, the
ordering occurring when $u$ increases is characterized by larger
values of $\langle m \rangle$ in an intermediate temperature
range. As $v/u$ increases, the maximal value of $P_{-}$ gets closer 
to $1/2$, which is the value found in the ordered phase discussed
previously in the isotropic case. Notice that the extent of the
temperature range where $P_{-}$ approaches $1/2$ increases with
the length $L_{\tau}$ of the system (see $v/u=0.6$ curves in the 
Fig.~\ref{manis}), which supports the scenario
of order by disorder: the $P_{-}=1/2$ asymptotic value can be
understood by the number $O(2^{L_{\tau}})$ of lowest-energy
defects (which are lowest-energy configurations in the isotropic
case). With interaction anisotropy, these defects are allowed only
at finite temperatures, and proliferate down to lower temperatures
as $u-v$ decreases. This clearly shows that the ordered phase in the isotropic 
triangular lattice model differs both qualitatively and quantitatively (by 
the number of lowest-energy configurations) from that of its anisotropic
version, and is a good example of the ``order by disorder'' scenario.

At this stage, we have three indications (from the correlation
length, the specific heat and $\langle m \rangle$) that the
isotropic triangular interacting dimer model could order at low
temperatures. In all cases, the transition temperature was seen to
decrease quite consequently with system size. To check whether the
system could order at finite $T$ in the thermodynamic limit, we
perform a finite-size scaling of these effective transition
temperatures. In the insert of Fig.~\ref{m1m2}, the finite-size
temperature transitions $T_{c}^{1}(L)$ and $T_{c}^{2}(L)$  corresponding to 
respectively the maxima of the specific heat and of $d\langle m \rangle /du$, 
are plotted as a function of inverse transverse system size $1/L$. A precise
finite-size scaling form is difficult to determine (due noticeably
to large error bars), but all reasonable finite size
dependence ({\it e.g.} linear or quadratic in $1/L$) lead to a
finite value of $T_{c}$ in the thermodynamic limit. A rough
estimate can be made with the help of linear interpolation (dashed line on
the figure) and gives $T_{c}=0.2 \pm 0.05$ (in units of $u$).

Our numerical results therefore seem to be consistent with a
finite temperature ordering of the isotropic lattice model,
probably triggered by an order by disorder
mechanism~\cite{Villain,SH}. The sharp behavior of both specific
heat and ``order parameter'' at the transition (see
Fig.~\ref{m1m2}), as well as the absence of any criticality behavior
in the central charge, suggests that this transition is first order. We
finish by noting that the samples used in the computations are of
relatively moderate size, and that it is still possible that the
extrapolated transition temperature actually vanishes with larger
samples available. To have a better understanding of this
transition with the presently available system sizes, one could
also study this model on samples with other geometries and
adapted order parameters, and check if extrapolations of
finite-size temperatures on different geometries give the same
$T_{c}$ in the thermodynamic limit. Another insight could be given
by Monte Carlo simulations for this model. We expect however these
simulations to be difficult as the presence of the low-energy
modes (and the corresponding large degeneracy of the ground-state)
will certainly induce ergodicity and freezing problems in the
Monte Carlo process~\cite{Castelnovo03}. Such investigations are beyond the
scope of the present paper.


\section{\label{sec: Conclusions}
Conclusions}

To summarize, we have constructed a simple classical interacting dimer model
on a lattice that interpolates between square and
triangular lattices. This is of particular interest since the two
limiting models singularly exhibit very different behaviors at
infinite temperature (critical and short-ranged phases for the square and
triangular lattices respectively). Since the topology of the triangular lattice can be
simply obtained from the square lattice by adding one extra
diagonal bond on each square plaquette, we have introduced a fugacity
parameter for these extra bonds. Similarly, the extra bonds leading to new
local interactions in diamond-like plaquettes, we have also
considered dimer interactions on these plaquettes, differing from those on the
square plaquettes.

\begin{figure}
\begin{center}
\includegraphics[height=5cm]{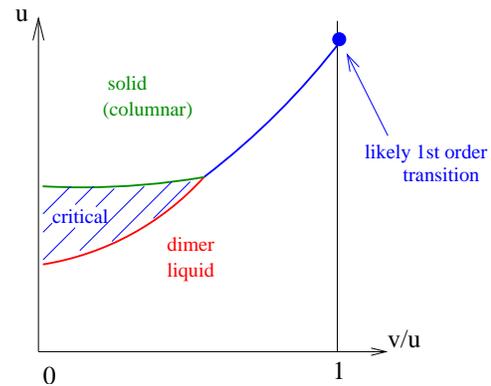}
\caption{\label{diagproj} Schematic projective phase diagram of the dimer model
on the triangular lattice ($z_{D}=1$, line E of Fig.~\ref{path}), in terms
of the interaction $u$ and interaction anisotropy $v/u$. }
\end{center}
\end{figure}

This anisotropic dimer model has been investigated in great details
using simple considerations as well as numerical transfer matrix techniques (on strips with up to 14 sites wide) supplemented
by predictions from conformal field theory. A very rich phase
diagram has been obtained with, in particular, a novel
intermediate behavior where the critical phase is now restricted
to a finite intermediate temperature range and
does not extend, as for the pure square lattice model, up to infinite
temperature. At high-temperature, the critical phase is hence replaced by a
liquid dimer phase as in the isotropic triangular lattice. It is
of interest that such a behavior appears immediately for arbitrary
small fugacity of diagonal bonds so that one can view the typical
behavior of the square lattice as a ``singular'' limit. However, we
note that the critical phase itself survives in an extended
vicinity of the square lattice model (although its extension in
temperature jumps abruptly) and this stability is
therefore {\it not} related to the bipartiteness of the lattice.
While the critical phase disappears progressively with increasing
interaction anisotropy, the ordered phase survives for any range
of anisotropy, as well as the high-T liquid dimer.
Criticality in this model hence requires to have sufficiently anisotropic
interactions. These findings are reproduced in the schematic phase
diagram of Fig.~\ref{diagproj} for the case of isotropic dimer fugacity
$z_D=1$. Lastly, we have devoted special attention to the
limiting case of the isotropic dimer model on the triangular
lattice where we found evidence of a direct (likely
first-order) transition between the high-T liquid phase and a
low-T ordered phase triggered by an order-by-disorder mechanism.

\begin{acknowledgments}
We thank F. Becca, S. Capponi and C. Castelnovo for fruitful discussions,
the {\it Agence Nationale de la Recherche} (France) for
support and IDRIS (Orsay, France) for computer time.
\end{acknowledgments}


\begin{appendix}
\section*{Transfer matrix analysis of the anisotropic
dimer model}

Other works~\cite{Alet0506,CCMP} already discussed the
construction of the TM for an interacting dimer model. We give a
few technical details related to the triangular lattice and to the observables studied in the body of the text.

\subsection{\label{conf} 1D and 2D dimer configurations}
The TM allows to treat a 2D $L \times L_{\tau}$ system with
periodicity in the transverse direction ($x$) as a
succession of $L_{\tau}$ 1D rows of lengths $L$. In a row at
abscissa $\tau$ (the $\tau$ axis being the direction of propagation
of the TM), a dimer configuration $C$ is specified by the dimer
occupation of horizontal (oriented along ${\bf e}_{x}$) links at
abscissa $\tau$ and of vertical and diagonal links connecting sites at
abscissa $\tau$ to sites at $\tau+1$, at relative positions ${\bf e}_{\tau}$
and  ${\bf e}_{\tau}-{\bf e}_{x}$ respectively. The number of such 
configurations, which is the size of the TM, is $3^L$ (This number can be 
evaluated as $\Tr(\bf{M}^L)$ where $\bf{M}$
is a \textit{one-dimensionnal transfer matrix} defined between the $4$ 
dimer configurations allowed on a triangle, taking into account the hard-core 
constraint).

\begin{table}
\begin{tabular} {|c|c|c|c|}
\hline
$L$ & $\dim({\bf T}^{e}(q=0))$ & $\dim({\bf T}^{e}(q=\pi))$ &
$\dim({\bf T}^{o}(q=0))$ \\
\hline
$8$ & $424$ & $421$ & $410$\\
\hline
$10$ & $2980$ & $2929$ & $2954$\\
\hline
$12$ & $22218$ & $22207$ & $22150$\\
\hline
$14$ & $170980$ & $170665$ & $170822$\\
\hline
\end{tabular}
\caption{\label{sizes} Dimensions of the TM restricted by
translation invariance to symmetry sectors of wave vectors $q=0$ and $q=\pi$,
and by topological invariance to \textit{even} and \textit{odd} sectors, for
system widths $L$ up to $14$.}
\end{table}

The topology of the triangular lattice is such that on a $L$-wide
cylinder, the parity of the number of vertical or diagonal dimers
linking two successive rows is conserved along the $\tau$ axis
(assuming that $L$ is even). This defines two topological sectors
(\textit{even} and \textit{odd}), and the TM is block-diagonal
with corresponding blocks $\textbf{T}^{e}$ and $\textbf{T}^{o}$
respectively. We also use the invariance by translation along the
$x$ axis to reduce the size of $\textbf{T}$. Translation
invariance allows to consider, instead of $3^{L}$ 1D
configurations, only their representatives in a symmetry sector
(defined by the transverse wave vector $q$), from which other
configurations are obtained by translations along ${\bf e}_{x}$.
Note that the TM elements take into account not only
representatives but also all configurations connected to a given
representative. These two invariances allow to reduce considerably
the TM size in various symmetry and topological sectors. We give
in Table~\ref{sizes} the size of the different sectors useful in
the following for $L$ up to $14$. Values of non-zero TM elements,
\textit{i.e.} between compatible 1D-configurations, depend on the
number of doubly occupied plaquettes of both types and of diagonal links
occupied in these configurations.

\subsection{\label{part} Partition function and correlation functions}

With the transfer matrix $\textbf{T}$ constructed, we have access to statistical
properties of the model on a $L \times L_{\tau}$ torus by means of the
partition function, which reads
$$
Z(u,v,z_{D})=\Tr({\bf T}^{L_{\tau}})=
\sum_{C_{1},C_{2}\ldots C_{L_{\tau}}} {\bf T}_{C_{1},C_{2}} \ldots
{\bf T}_{C_{L_{\tau}},C_{1}}
$$

{\it Internal energy and specific heat --- } From the TM elements, one also
gets the internal energy of the system:
\begin{eqnarray*}
\langle E \rangle =-\frac{1}{Z}\sum_{C_{1},C_{2} \ldots C_{L_{\tau}}} {\bf
  T}_{C_{1},C_{2}} \ldots {\bf T}_{C_{L_{\tau}},C_{1}} . \nn\\
 \ln({\bf T}_{C_{1},C_{2}} \ldots {\bf T}_{C_{L_{\tau}},C_{1}}).
\end{eqnarray*}

We computed the specific heat per site $c_v=\frac{1}{L.L_{\tau}}
d\langle E\rangle/dT$ by a numerical differentiation of $\langle E
\rangle$ with respect to the inverse of the coupling strength. For
the calculations of Sec.~\ref{isotri}, $\langle E \rangle$ was
computed exactly for $L=4,6,8$-wide systems (with $L_\tau \gg L$
so that $c_{v}$ does not depend on $L_{\tau}$). For wider systems
$L=10,12$, $\langle E \rangle$ was estimated by searching the
leading TM eigenvector and evaluating from it the energy per row
$\langle E \rangle/L_\tau$ (the consistency of this method, valid in the limit of infinitely long
systems, was checked for $L=8$ first).

{\it Correlation functions --- } Correlation functions are computed by means of
TM iterations. Dimer-dimer correlations, for two dimers with a specified
orientation and relative position $\textbf{r}$, are obtained as the ratio over
the partition function $Z$ of a modified partition function $Z'$ taking into account only the
2D-configurations where these two dimers are present. Mathematically,
a projector $\textbf{P}$ onto the corresponding 1D configurations is inserted
in the matrix product in $Z'$ at positions corresponding to both dimers:
\begin{eqnarray*}
C({\bf r})=\frac{Z'}{Z}=\frac{\Tr({\bf P}\,{\bf T}^{r_\tau}\,{\bf P}\,{\bf T}^{L_\tau-r_\tau}
)}{\Tr({\bf T}^{L_\tau})}
\end{eqnarray*}
where $r_\tau$ is the projection onto the $\tau$ axis of the relative
position $\bf{r}$. Even if we do not discuss them in the body of
the text, monomer-monomer correlations are also accessible via TM
calculations. In the TM language, the insertion of a monomer in a given row requires to replace the TM
by a modified matrix $\textbf{T'}$ which forbids the occupation of
links emerging from this site (this involves a change of
topological sector). For both dimer-dimer and monomer-monomer
correlations, we checked that in the non-interacting case, the exact results
of Ref.~\onlinecite{FMS02} are recovered.

\subsection{\label{corr} Analysis of real-space correlations}

When the system is in a critical phase, conformal
transformation techniques~\cite{Cardy} can be used to analyze the
real-space decay of correlation functions on a $L$-wide infinite cylinder.
Consider the correlation of two dimers (or monomers) located at a
relative distance $r_{\tau}$ along the cylinder axis. If
the exponent of the infinite-plane power-law correlations is
$2\alpha$, correlations on the cylinder scale as:
$$
C(r_\tau)\sim \Bigg(\cosh(\frac{2\pi r_{\tau}}{L})-1 \Bigg)^{-\alpha},
$$
which can be approximated, for distances much larger than
$L/(2\pi)$, by an exponential decay with a correlation length
$L/(2\pi \alpha)$. We checked our calculations on the
non-interacting case on the square lattice for a $L=6$ cylinder.
We found a correlation length $\xi=1.02(2)$, close to the expected
(thermodynamical limit) value  $L/(2\pi d_{1,0})=3/\pi\sim 0.95$.
The small discrepancy can be attributed to the $1/r^4$ correction
present in the correlation function~\cite{Fish-Steph}. Repeating
the same calculations for monomer correlations, we
extracted from the exponential decay a critical exponent
$d_{0,1}\sim 0.26$, in good agreement (for such a small $L$ value)
with the exact value $1/4$.

\subsection{\label{CFT} CFT analysis of largest TM eigenvalues}

The CFT analysis is also useful to compute the central charge $c$,
as well as  the scaling dimension of the dimer $d_{1,0}$ and
monomer $d_{0,1}$ operators, from the size dependence of the
leading TM eigenvalues~\cite{Alet0506}. The central charge is estimated by the $L$-dependence of the
largest eigenvalue $\Lambda_{0}$ of $\textbf{T}$, which is
directly connected to the free energy per site $f_{0}$, in the
limit of an infinitely long system: \beq f_{0}=-\frac {1}{L}
\ln(\Lambda_{0})=f^{*}-\frac{\pi c}{6 L^{2}} +
o(\frac{1}{L^{2}})\, .\nn \eeq In practice, we add a
$\frac{1}{L^{4}}$ term to the previous fitting
expression~\cite{CCMP}, which is justified by symmetry reasons and
improves the agreement between raw data and the fitted expression
of $f_{0}$. Estimating $c$ thus requires to perform a fit of
$f_{0}$ with at least 3 values of $L$. The size of the TM
is a limitation to the number of sizes accessible; nevertheless,
one can use the already mentioned topological and translation
invariance to adress larger sizes. Indeed, one can assert by
symmetry reasons and check numerically that the leading eigenvalue
of the TM is found in the $q=0$ \textit{even} sector. The scaling dimensions of dimer and
monomer exponents are determined from the largest eigenvalues in other symmetry sectors~\cite{Alet0506}:
\begin{eqnarray*}
-\frac{1}{L} \ln\Big(\frac{\Lambda_{0}^{e}(q=\pi)}{\Lambda_{0}^{e}(q=0)}\Big)
&=&\frac{2 \pi d_{1,0}}{L^2} + O(\frac{1}{L^4}) \nn \\
-\frac{1}{L} \ln\Bigg(\frac{\Lambda_{0}^{o}(q=0)}{\Lambda_{0}^{e}(q=0)}\Bigg)
&=&\frac{2 \pi d_{0,1}}{L^2} + O(\frac{1}{L^{4}}).
\end{eqnarray*}
This allows to determine $c$ and $d_{e,m}$ (with $(e+m=1)$) from
the power method~\cite{Wilkinson}, requiring less memory than for a
full diagonalization. To minimize finite-size effects on $c$ and $d_{e,m}$, we use different
estimates for these quantities. Each estimate is the result of a fit with
either $3$ (for $c$) or $2$ (for $d_{e,m}$) consecutive values of $L/2$.
In this way one obtains estimates $c(L-4,L)$  and $d_{e,m}(L-2,L)$ (for $L \le
14$). The finite-size effects on these estimates become smaller when increasing
$L$; in order to obtain more reliable values we make extrapolations
of these estimates, assuming that they scale as~\cite{CCMP}:
\begin{eqnarray*}
c(L-4,L) & \sim & c^{*} + \frac{K}{(L-4)^2}\nn\\
d_{e,m}(L-2,L) & \sim & d_{e,m}^{*} + \frac{K'}{(L-2)^2}
\end{eqnarray*}
Such estimates and their extrapolations are displayed in
Fig.~\ref{plat0}, \ref{plat1} and \ref{plat2}. In general, the
$c^*$ extrapolations were done with estimates $c(8,12)$ and
$c(10,14)$, while for $d_{e,m}^{*}$ extrapolations we used the
estimates $d_{e,m}(8,10)$, $d_{e,m}(10,12)$ and $d_{e,m}(12,14)$.
The difference between the extrapolation and the closest estimate
(the one with the largest $L$) gives the order of magnitude of
the error.

\end{appendix}

\end{document}